\documentclass[draft]{agujournal2019}

\usepackage{url} 
\usepackage{lineno}
\usepackage[inline]{trackchanges} 
\usepackage{soul}
\usepackage{algorithm}
\usepackage{algpseudocode}
\usepackage{amsmath}

\raggedright
\draftfalse

\journalname{Space Weather}
\hypersetup{final}

\begin{document}
%
%


\title{Solar Transient Recognition Using Deep Learning (STRUDL) for heliospheric imager data}

%
%




\authors{M. Bauer\affil{1,2}, J. Le Louëdec\affil{1}, T. Amerstorfer\affil{1}, L. Barnard\affil{3}, D. Barnes\affil{4}, H. Lammer\affil{5}}


\affiliation{1}{Austrian Space Weather Office, GeoSphere Austria, Graz, Austria}
\affiliation{2}{Institute of Physics, University of Graz, Graz, Austria}
\affiliation{3}{Department of Meteorology, University of Reading, UK}
\affiliation{4}{RAL Space, STFC Rutherford Appleton Laboratory, Didcot, UK}
\affiliation{5}{Austrian Academy of Sciences, Space Research Institute, Graz, Austria}





\correspondingauthor{Maike Bauer}{maike.bauer@geosphere.at}



\begin{keypoints}
\item We developed a machine learning model to automatically detect and track coronal mass ejections in heliospheric images
\item Automatic detection provides internally consistent results, helping to address some of the challenges of manual tracking
\item The model performs well on clear events but struggles with faint or overlapping structures, highlighting areas for future work
\end{keypoints}

%
%

%
%


\begin{abstract}

Coronal Mass Ejections (CMEs) are space weather phenomena capable of causing significant disruptions to both space- and ground-based infrastructure. The timely and accurate detection and prediction of CMEs is a crucial steps towards implementing strategies to minimize the impacts of such events. CMEs are commonly observed using coronagraphs and heliospheric imagers (HIs), with some forecasting methods relying on manually tracking CMEs across successive images in order to provide an estimate of their arrival time and speed. This process is time-consuming and results may exhibiting considerable interpersonal variation.

We investigate the application of machine learning (ML) techniques to the problem of automated CME detection, focusing on data from the HI instruments aboard the STEREO spacecraft. HI data facilitates the tracking of CMEs through interplanetary space, providing valuable information on their evolution. Building on advances in image segmentation, we present the Solar Transient Recognition Using Deep Learning (STRUDL) model. STRUDL is designed to automatically detect and segment CME fronts in HI data. We address the challenges inherent to this task and evaluate the model’s performance across a range of solar activity conditions. To complement segmentation, we implement a basic tracking algorithm that links CME detections across successive frames, thus allowing us to automatically generate time-distance profiles.

Our results demonstrate the feasibility of applying ML-based segmentation techniques to HI data, while highlighting areas for future improvement, particularly regarding the accurate segmentation and tracking of faint and interacting CMEs.

\end{abstract}

\section*{Plain Language Summary}

Coronal Mass Ejections (CMEs) are large bursts of solar material that can disrupt satellites, power grids, and communication systems on Earth. To prepare for these events, CMEs need to be detected and assessed quickly. This is often done by hand, but the interpretation of data is not always straightforward and results can vary between forecasters. In this study, we present a machine learning model called Solar Transient Recognition Using Deep Learning (STRUDL) to automatically detect and track CMEs in images from NASA’s STEREO spacecraft. Our model can follow CMEs as they move away from the Sun, which helps us understand how they evolve over time. While our model works well for clear events, it struggles with faint or overlapping CMEs. This work demonstrates that machine learning can be a valuable tool for monitoring CMEs, although further improvements are necessary to effectively handle complex events.

%
%

%


%
%
%
%

\section{Introduction}\label{section:Introduction}

Coronal mass ejections (CMEs) are explosive eruptions of plasma and magnetic field from the solar corona. Their speed can range from 100 km/s at the lower end to over 2500 km/s for fast CMEs \cite{yashiroCatalog2004,manoharanCoronal2011}. Depending on the CME's speed and magnetic field orientation, it may have a significant impact on planets and their magnetospheres. CMEs pose a potential risk to astronauts in space, satellites, and power grids on Earth, making identification and arrival forecasting of Earth-directed CMEs crucial for risk mitigation.

The study of CMEs dates back to the early 1970s, when space-based coronagraphs aboard the OSO-7 \cite{koomenWhite1975} and Skylab \cite{macqueenPhotography1975} missions captured the first images of CMEs. The Solar Mass Ejection Imager \cite<SMEI;>[]{jacksonSolar1997}, launched in 2003 aboard the Coriolis spacecraft, was designed to investigate CMEs as drivers of space weather, specifically to assess the viability of using image data to track CMEs and forecast their arrival at Earth \cite{webbSolar2002}. A major advancement in our understanding of CMEs came with the launch of the twin Solar TErrestrial RElations Observatory \cite<STEREO;>[]{kaiserSTEREO2008} spacecraft in 2006. Consisting of STEREO-A and STEREO-B, the mission provided a stereoscopic view of the Sun, facilitated by two coronagraphs and two wide-field heliospheric imager \cite<HI;>[]{eylesHeliospheric2009} instruments on board each spacecraft. The combined field-of-view (FOV) of the HI instruments extends from 4 to 88$^{\circ}$ of elongation from Sun-center, allowing for continuous observation of CMEs as they move away from the Sun. Since the start of the STEREO mission, more spacecraft equipped with HI instruments have been launched \cite<e.g. PSP/WISPR, SoloHI, PUNCH;>[]{vourlidasWideField2016,howardSolar2020,deforestPolarimeter2022}. These missions further aid the analysis of CME structure and kinematics by providing complementary observations from different viewpoints.

To enable the study of CMEs and their properties, a number of catalogs utilizing different identification techniques and instruments have been published over the years. \citeA{barnardSolar2014} introduced a catalog based on the Solar Stormwatch I citizen science initiative, where volunteers were asked to analyze data from the STEREO/HI instruments to identify and track CMEs. By aggregating multiple user observations, time-elongation profiles could be generated for each event. The HICAT catalog, first presented by \citeA{harrisonCMEs2018} and developed under the EU FP7 HELCATS project, is another, more recent example. It comprises over 3000 CMEs identified through manual inspection of STEREO/HI data. To account for ambiguity, each event is categorized as either good, fair, or poor in quality, depending on how certain the identification is. An event is classified as good if any person experienced in working with STEREO/HI data would consider it a CME. Conversely, an event is classified as poor if it is likely to be a CME, but there is considerable uncertainty due to factors such as brightness variation, topology, or the presence of data gaps. A fair quality event lies between the other two definitions, meaning that the CME is relatively clearly visible, but not completely unambiguous. The catalog was updated by \citeA{barnesHELCATS2018} to include geometric parameters for a subset of the CMEs in HICAT. Furthermore, the updated catalog, termed HIGeoCAT, provides time-elongation profiles along the position angle (PA) that corresponds to the center of the angular extent of each CME. Both HICAT and HIGeoCAT start in April 2007, the beginning of the official science-phase for the SECCHI suite of instruments, and are continuously updated. All of these catalogs, while valuable, are time-consuming in their creation and not updated in real-time.

In recent years, several automated CME detection algorithms have been developed to address some of the issues that manual methods face. Many of these algorithms are meant for use with coronagraph data \cite<e.g. CACTus, CAMEL;>[]{robbrechtAutomated2004,wangNew2019}. Automated detection in HI data has received somewhat less widespread attention, likely due to the fact that CMEs are more difficult to identify in HI compared to coronagraph data. A CME catalog based on SMEI data was published by \citeA{tappinAutonomous2012}, though they reported a high number of false positive detections due to the noisy nature of the data. \citeA{barnardDifferences2015} developed the J-Tracker method, which relies on Canny edge detection to extract transients from STEREO/HI time-elongation maps, in which the CME is viewed along one particular PA. More recently, CACTus, which was originally developed to automatically detect CMEs in images from the C2 and C3 Large Angle Spectrometric Coronagraph \cite<LASCO;>[]{bruecknerLarge1995} instruments aboard the Solar and Heliospheric Observatory \cite<SOHO;>[]{domingoSOHO1995}, has been adapted to work on time-elongation maps of STEREO/HI data \cite{pantAutomated2016}. It utilizes image processing techniques to identify CME fronts based on predefined criteria, aiming to minimize the impact of human subjectivity on detection. \citeA{rodriguezComparing2022} applied the modified CACTus algorithm to a range of years, spanning from April 2007 to August 2020, and found that, compared to the manually created HICAT catalog, CACTus tends to overestimate the number of CMEs, likely due to the algorithm frequently categorizing parts of the same front as separate CMEs. 

Moving away from efforts focused on tracking CMEs in time-elongation maps, \citeA{kirnosovCombining2016} developed an automated tracking algorithm that utilizes images from both the COR2 and HI1 instruments aboard STEREO-A. Their algorithm yielded satisfactory estimates of speed and PA for the 15 events under study, but has not been tested on a larger dataset. An effort to manually track CME fronts across multiple PAs in a large number of STEREO/HI images was made by the Solar Stormwatch II project \cite<SSW-II;>[]{barnardTesting2017}. Designed as a follow-up study to Solar Stormwatch I, SSW-II was a citizen science project seeking to improve space weather forecasting by recruiting volunteers to track CMEs in STEREO/HI images. Citizen scientists analyzed a number of events that occurred in the 2011 -- 2012 time period, thus generating a valuable source of data cataloging solar storms in STEREO/HI data. \citeA{barnardTesting2017} posit that tracking in HI data directly may have several advantages compared to tracking done in time-elongation maps. These include a more complete picture of the complex structure of the front, as well as eliminating uncertainty in the time coordinate resulting from the averaging and interpolation conventionally applied to time-elongation maps.

The detection of CMEs in image data shares many similarities to problems encountered in other disciplines, such as biomedical imaging. Recent advancements in machine learning, particularly deep learning, offer promising alternatives for automating the detection of solar phenomena in observational data. In particular, convolutional neural networks \cite<CNNs;>[]{lecunConvolutional2010} have demonstrated success in various image recognition tasks. Among these, the U-Net \cite{ronnebergerUNet2015} and its variants, such as the ResUNet++ \cite{jhaResUNet2019}, have proven effective in a variety of contexts \cite{cuiPulmonary2019,wuAutomatical2019,fangRCAUNet2019}. In recent years, machine learning techniques have started gaining increasing significance in the space sciences, yielding promising results across different tasks covering domains ranging from the solar surface \cite{jarolimAdvancing2024}, to interplanetary space \cite{rudisserAutomatic2022} all the way to the terrestrial atmosphere \cite{malikTransient2023} and beyond \cite{johnsonRotNet2020}.

In this work, we present an automatic detection pipeline for CMEs based on a U-Net model, specifically tailored to segment CME fronts in STEREO/HI data, which we term STRUDL. We also introduce an automatic tracking algorithm designed to work with the output that STRUDL provides. In Section \ref{sec:data} of this article, we describe the data products used to train, validate, and test STRUDL with emphasis on how we incorporated data from the SSW-II project into our model. We also give an overview of the methods to which we compare our model. In Section \ref{sec:methods}, we provide an in-depth description of our model architecture, as well as the post-processing of results and the techniques used to evaluate the model's performance. In Section \ref{sec:results}, we summarize the results.

\section{Data}
\label{sec:data}

We conducted our experiments utilizing data from the HI1 instrument aboard the STEREO-A spacecraft. The dataset comprises a total of 13876 $1024 \times 1024$ images from the time period between January and August 2010, as well as January to May 2012. We identify a total of 258 CMEs in the given time frames. During these periods, the 11-year solar cycle emerged from a minimum and started moving towards its peak in early 2014. Choosing these time periods ensures that the dataset covers non-interacting (more common around solar minimum) as well as interacting (more common around solar maximum) events \cite{rodriguezgomezClustering2020} while also avoiding the latter part of the cycle in which STEREO-A moved towards the back of the Sun, placing it in an increasingly unfavorable viewing position.

The STEREO/HI data are preprocessed to enhance the visibility of CMEs and suppress noise. The IDL SolarSoft routine \texttt{secchi\_prep.pro} is commonly used for this purpose. In this work, we rely on the Python HI-processing suite (see Section \ref{sec:openresearch}), which applies the same methods as \texttt{secchi\_prep.pro}. These include the masking of saturated columns, often caused by planets passing through the spacecraft's FOV, correction of image-smearing occurring due to the shutter-less nature of the imagers' cameras, and the application of flat-field, distortion, and vignetting calibrations. To enhance CME visibility, we make use of running difference images. These data products are created by subtracting the previous frame from the current one, yielding a final image in which the dynamic structure of the CME are amplified compared to the static background.

The lack of a machine learning-ready dataset containing STEREO/HI data required manual annotation of images from the chosen time period. For the 2010 time period, the CMEs appearance in HI was identified according to the HICAT catalog where applicable, and the perceived front was annotated in each image by expert annotators. A 30-pixel wide region was chosen to mark the area of interest. The marked region later undergoes two iterations of morphological dilation with a disk-shaped kernel of size 2 to better capture the uncertainty inherent to the task. The combination of linewidth and post-processing approach was selected based on visual inspection across the dataset and was found to consistently encompass the area of interest along the CME front.

\begin{figure}
\includegraphics[width=\textwidth]{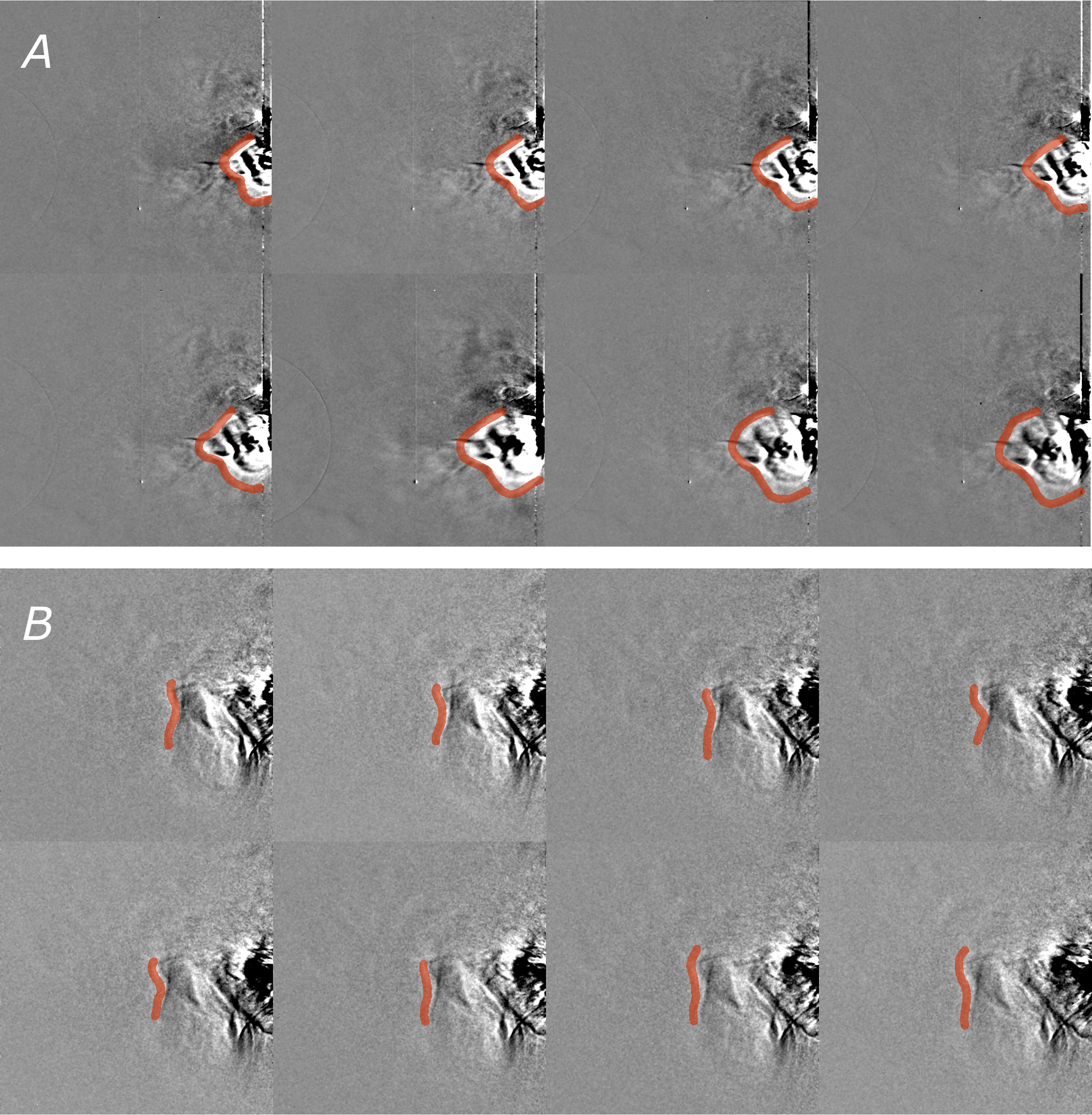}
\caption{Sequences of STEREO-A HI1 images from January 2010 (A) and May 2010 (B). The data have been post-processed and subsequently reprocessed into running difference images. The manually annotated CME front is overlaid in red.}
\label{fig:stereo_data}
\end{figure}

For the 2011 time period, the data collected by the SSW-II project served as a starting point for the annotation. The objective of SSW-II was to track the leading edge of CMEs propagating through STEREO-A and STEREO-B HI1 images. For the period of 2011--2012, running difference images of CMEs, referred to as ``assets'', were produced to create the data set that the citizen scientists would use to characterize the CMEs. The HICAT catalog was used to identify time windows when a CME would likely be in either the HI1A or HI1B field of view. For the 2011--2012 period, there are 307 entries in the HICAT catalog. The assets were presented as ``subjects'', where a subject refers to a grouping of three consecutive images. A training exercise was provided to demonstrate to the citizen scientists a range of appearances that CME fronts can take in HI1 data. Participants would then draw up to seven polygons on the three assets in each subject, tracking any feature they identify as a CME front. Analyzing three images at a time allowed participants to gauge the evolution of the CME between frames and was found to aid CME identification. Subjects were constructed so that they overlapped the previous/next subject by one asset, which enabled the continuous tracking of CME fronts. In total, this required 6672 subjects, which were analyzed by 9107 participants, resulting in 207218 classifications of CME fronts.

\begin{figure}
\includegraphics[width=\textwidth]{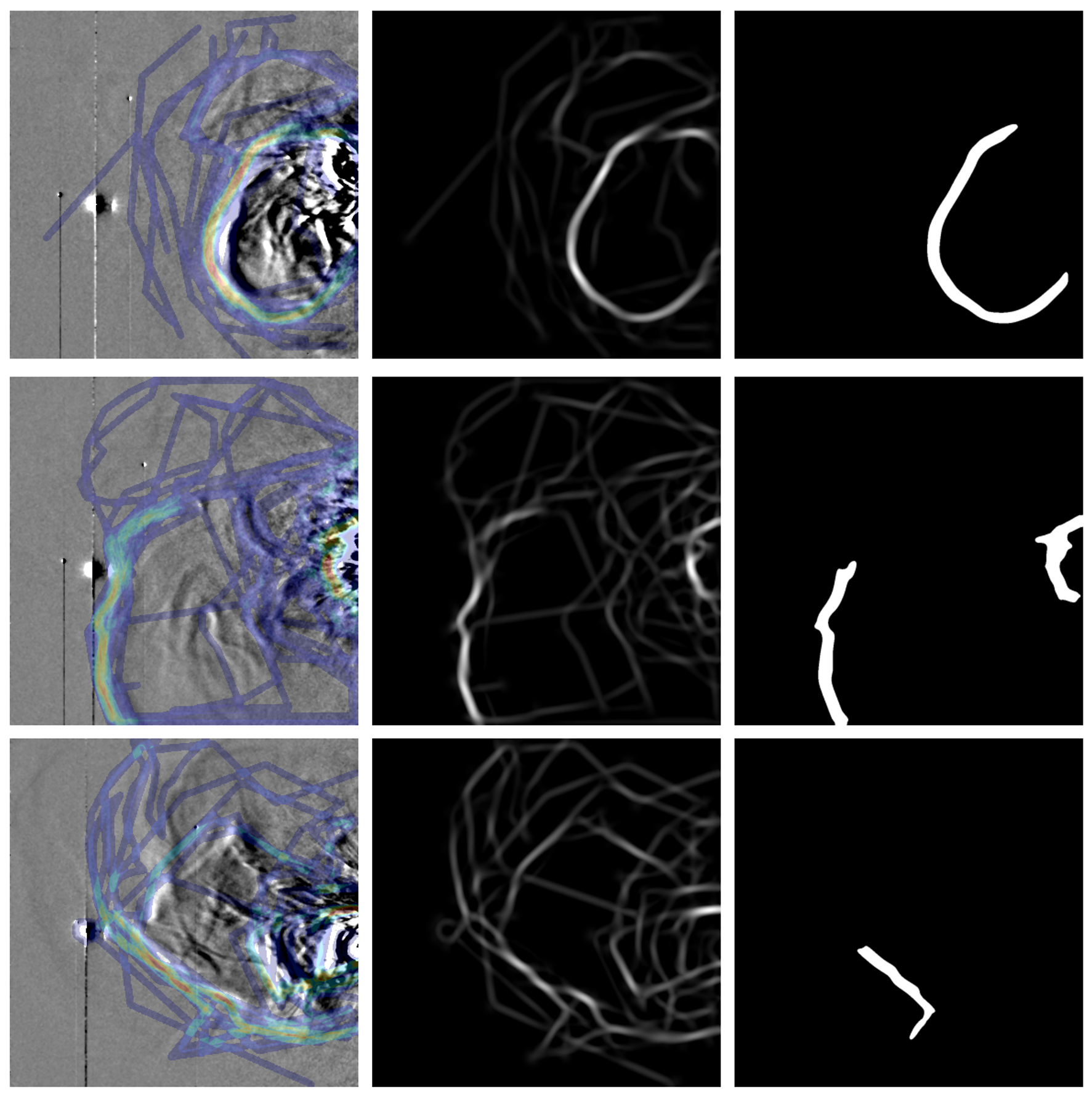}
\caption{Example of post-processing steps to obtain consensus masks from the Solar Stormwatch II citizen science project, which we use as a source of additional data to train, validate, and test STRUDL. The leftmost column shows the summed annotations plotted over the corresponding STEREO/HI running difference image. The middle column shows the same annotations after application of a Gaussian filter with $\sigma = 8$, followed by a Sato filter with $\sigma \in [1,10]$. In the rightmost column, the final binary consensus masks, obtained after normalizing the image and applying a threshold $t = 0.3$, are shown.}
\label{fig:ssw_data_reduction}
\end{figure}

To incorporate SSW-II data into our dataset, the various annotations by different participants must be combined into a single mask for each image. To reach a consensus, a series of processing steps is applied to the annotations of each image. As a first step, all annotations are summed up, followed by the application of a Gaussian filter. A Sato filter is used to extract curves from the summed masks, and the final binary mask is obtained by applying a threshold. Regions of interest are extracted from the binary mask and subsequently separated into distinct CMEs with separate labels. The processing steps were determined through iterative experimentation guided by visual inspection; examples of the masks before and after undergoing processing can be seen in Figure \ref{fig:ssw_data_reduction}.

\section{Methods}
\label{sec:methods}

\subsection{Semantic Segmentation}

The model described in the following section was implemented using the PyTorch framework, an open-source Python library. For access to the code used in this paper, as well as to the pre-trained models, please refer to Section \ref{sec:openresearch}.

We compile our data into non-overlapping sequences, each containing a consecutive series of 16 images. We aim to maintain a consistent ratio of CME to non-CME images in each sequence, although the ratio may be slightly higher or lower at times due to the constraint that sequences must not overlap. We use a stride of two when generating the input sequences, resulting in each frame appearing in multiple sequences. For the segmentation task, we employ a simple U-Net, adapted to work on three-dimensional data, the third dimension being time.

\begin{figure}
\includegraphics[width=\textwidth]{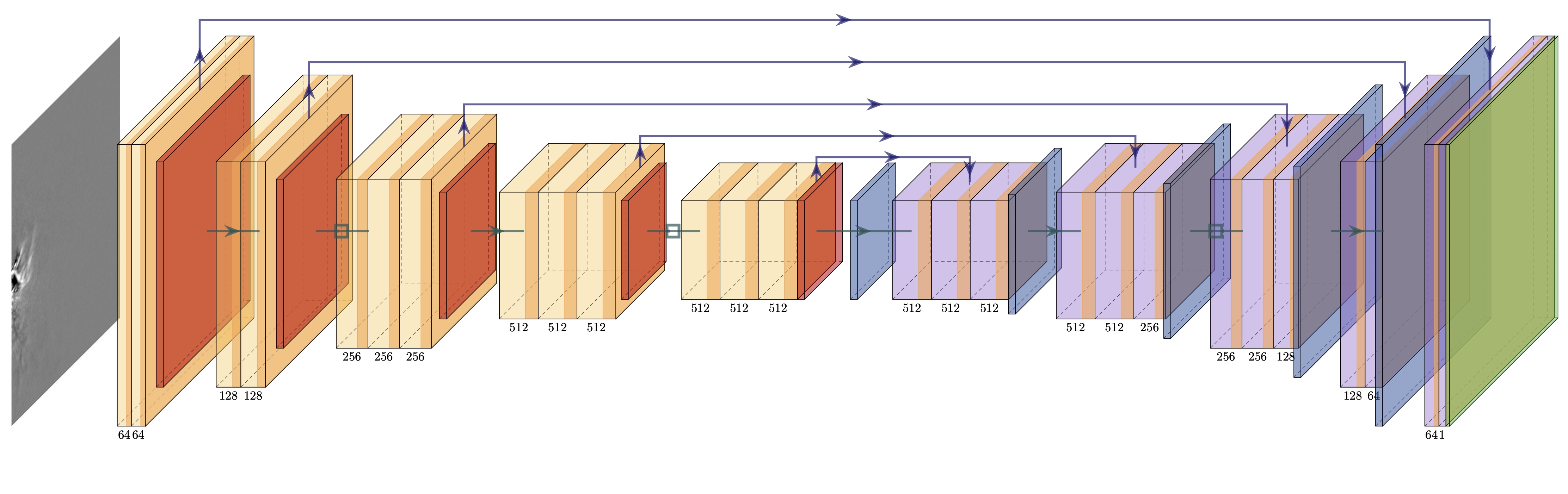}
\caption{Schematic illustration of our 3D UNet. The diagram illustrates the sequence of operations applied to the input data as it propagates through the network. The left-hand side of the network corresponds to the encoder path, the right-hand side to the decoder path. Arrows indicate the direction of the flow of the data. Some arrows are marked with a square, showing the presence of a dropout layer. Light orange blocks represent convolutions, while darker orange bands represent the ReLU activation function. The numbers below the convolutions indicate the size of the feature space after the convolution operation. Red and blue blocks indicate 3D max pooling and unpooling operations, respectively. Purple blocks indicate transpose convolutions. The terminal layer (green) normalizes the network's output between 0 and 1 using a sigmoid function. The purple arrows between the encoder and decoder paths of the network show the skip connections.}
\label{fig:UNET}
\end{figure}

An overview of our model architecture is given in Figure \ref{fig:UNET}. Our model consists of two separate paths, referred to as encoder and decoder. Given that our model is based on a U-Net architecture, these paths are symmetrical. The model's encoder path consists of five separate stages, each containing either two or three three-dimensional convolutional layers. Convolutional layers are the basic building blocks of a CNN and are designed to extract relevant features from the data. Each convolutional layer is followed by a so-called activation function. These functions transform their input non-linearly to allow the model to learn complex patterns. We chose the commonly used rectified linear unit (ReLU) activation function, given by $f(x) = max(0,x)$, for this purpose.

After passing through a group of convolutional layers along the encoder path, the output is spatially downsampled using three-dimensional max pooling layers, preserving important features while reducing the amount of time needed to process the data in subsequent layers. Conversely, our model’s decoder reconstructs the data’s original spatial resolution by progressively upsampling the feature maps using three-dimensional max unpooling layers. To obtain output data with the same shape as the input data, we apply transpose three-dimensional convolutions along the decoder path, effectively performing the inverse operations of the convolutional layers in the encoder path. Finally, we apply a sigmoid layer to our output. The sigmoid layer ensures that every pixel in our segmentation mask has a value between 0 and 1, giving us a probability that the pixel is part of a CME. The encoder and decoder stages of the model are linked by so-called skip-connections. These connections allow the output of a particular encoder level to be fed directly to the corresponding decoder level, instead of having to traverse the entirety of the network. This helps information to propagate more freely throughout the network and can improve segmentation results.

The model outputs segmentation masks for each individual frame within a sequence, providing pixel-wise labeling of CME fronts. We post-process the segmentation masks to fill small holes within continuous fronts and remove small objects likely to represent spurious detections. To improve the robustness of the model and reduce the risk of overfitting to the training data, we apply simple data augmentation techniques. Specifically, we randomly flip input images horizontally and vertically during training. This encourages the model to learn features that are invariant to orientation, thereby improving its ability to generalize to new, unseen data. Such augmentation techniques are widely used in computer vision tasks, and their value for learning invariance has been demonstrated numerous times \cite<e.g.>[]{mikolajczykData2018}. Furthermore, we include dropout layers in our model. These layers are designed to randomly ignore, or drop, parts of the input during training, effectively making the model more robust and further reducing the risk of overfitting to the training data.

To update our model’s parameters during training and guide it toward convergence, an optimization algorithm must be chosen. We use the Adam optimizer with an initial learning rate of $1 \times 10^{-5}$. Adam is a widely used, efficient method that adaptively adjusts the learning rate for each parameter individually, rather than relying on a single global learning rate. Training runs for 150 epochs, but only the model with the highest Intersection-Over-Union (IoU) score on the validation set is saved to prevent overfitting. The IoU score will be explained in detail in Section \ref{subsec:evaluation}.

We have to consider that pixels belonging to a CME represent only a small fraction of pixels in each image, if a CME is present at all, while background pixels dominate. This imbalance could cause the model to prioritize correctly identifying the background pixels over identifying the CME itself. To mitigate the impact of this issue, we select a loss function for training our model that is specifically designed for imbalanced datasets such as ours. In machine learning, the loss function measures how well a prediction matches the ground truth data, guiding the model's learning process. Binary Cross Entropy (BCE) is a loss function that's commonly used for classification tasks. It computes the loss for every pixel based on how different its predicted probability is from its true label. The larger the difference between the prediction and the true label, the larger the BCE loss for that pixel becomes. The Dice loss, on the other hand, focuses on the overlap between predicted and ground truth regions, making it useful for segmentation tasks. During training, we employ the asymmetric unified focal loss proposed by \citeA{yeungUnified2022}, which builds on both the well-established BCE and Dice losses while introducing additional terms to emphasize the importance of CME pixels. The model is trained on an NVIDIA GTX 4090 (24 GB) GPU with a batch size of 4, and training batches are randomly shuffled at the start of each epoch.

\subsection{Tracking}

To extract individual CME regions from the final segmentation images and track them across frames, we use algorithms implemented in Python’s scikit-image library. First, we segment the image into distinct regions using the connected component labeling algorithm, following the methods outlined in \citeA{fiorioTwo1996} and \citeA{wuOptimizing2005}. For tracking purposes, we extract the CME front by applying the skeletonize algorithm \cite{zhangFast1984} to the segmented area, yielding a line of pixel coordinates. We transform the pixel coordinates into world coordinates, which are the heliocentric latitude and longitude in the case of STEREO/HI observations. We use the equations found in \citeA{thompsonCoordinate2006} to transform these coordinates into elongation $\epsilon$ and position angle, as these are commonly used when tracking CMEs in HI data. 

We then employ a multi-step algorithm designed to associate CME fronts across successive images. An overview of the tracking algorithm, presented in pseudo-code, is provided in Algorithm \ref{alg:tracking}. First, all CME fronts detected within a single image are compared to each other. If their mean elongation values differ by less than $3^\circ$, they are assumed to belong to the same front and are merged. This threshold was chosen based on the effects of our segmentation pre-processing. Specifically, we apply morphological dilation to the masks to ensure that CME fronts are fully captured and not too narrowly defined. As a result, segmented regions can span several degrees in elongation, even when the actual CME front is more localized. The $3^\circ$ threshold accounts for this spatial spread, and ensures that fragmented detections of the same structure within a frame are grouped before further analysis. Next, each front is evaluated to determine whether it could represent the start of a new CME. This is based on its mean $\epsilon$. If a front appears at an elongation less than $7.3^\circ$, it is considered a potential new CME. If it appears at a greater distance, it is not considered a potential starting point, but is kept for further analysis. Once candidate CME fronts are identified in the current image, we attempt to associate them with CMEs in the previous frame, if there are any. Each front in the current image is compared to all fronts in the previous image using the difference in mean elongation $\Delta\epsilon$ at each PA. If $\Delta\epsilon$ lies within the threshold of $-3.2^\circ$ to $+6.7^\circ$, the fronts are considered to belong to the same CME.

If a front in the current image is not associated with any front from the previous frame but was flagged as a potential new CME, it is treated as the beginning of a new CME. Otherwise, it is discarded. In cases where the current image contains no CME candidates but the previous one did, we interpret this as the end of the tracked CME. If neither image contains any CME candidates, no action is necessary. To avoid large temporal gaps between images, the tracking algorithm is only applied when the time interval between two consecutive frames is less than 140 minutes. We discard CME tracks with maximum elongations below $12.9^\circ$, as such detections are unlikely to represent true CME events. The thresholds for identifying new fronts, associating them across successive images, and determining whether a tracked structure qualifies as a CME were established empirically. To define reasonable baseline values, we analyzed 255 CMEs listed in the HIGeoCAT catalog for the years 2021 and 2022. These years were chosen because they represent a range of solar conditions, and don't intersect with any of the data in our training, test, or validation sets.

We determined the 1st percentile of the maximum elongation reached to set the minimum elongation criterion for a CME, yielding $12.9^\circ$. To avoid misidentifying background features as CME origins, we used the 99th percentile of starting elongation values to define an upper bound of $7.3^\circ$ for new CME fronts. To derive the thresholds used for associating fronts between images, we evaluated the range of elongation differences between time steps for each of the 255 events. The 99th percentile of maximum elongation change was $3.7^\circ$, and the 1st percentile of minimum change was $-0.2^\circ$. Because our fronts represent broad structures rather than sharply defined edges, and we rely on mean elongation differences to compare them across images, we added a buffer of $3^\circ$ to these percentiles. This adjustment results in final association thresholds of $-3.2^\circ$ to $+6.7^\circ$.

\begin{algorithm}
\caption{Pseudo-code outlining the basic principles of our CME tracking algorithm.}\label{alg:tracking}
\begin{algorithmic}[1]
\Require $t_{i} - t_{i-1} < 140$

\Function{AGGREGATE FRONTS}{$front_i$}
    \ForAll{$front_i^n$ in $image_i$}
        \ForAll{$front_i^m$ in $image_i$}
            \If{$|$mean($\epsilon_i^n$) - mean($\epsilon_i^m$)$|<$ 3 }
                \State $front_i^n$ and $front_i^m$ belong to same CME
            \EndIf
        \EndFor
    \EndFor
\EndFunction

\hfill

\Function{IDENTIFY NEW CME}{$front_i^n$}
    \If{mean($\epsilon_i^n$) $<$ 7.3}
        \State $front_i^n$ is potential start of new CME
    \Else
        \State $front_i^n$ is not start of new CME
    \EndIf
\EndFunction

\hfill
    \State AGGREGATE FRONTS($front_{i}$)\Comment{Combine fronts in same image}    
    \ForAll{$front_i^n$ in $image_i$} \Comment{Check which front could be start of new CME}
        \State IDENTIFY NEW CME($front_i^n$)
    \EndFor

    \If{count($front_{i}$) $> 0$ and count($front_{i-1}$) $>$ 0} \Comment{Check association with CMEs in previous image}
        \ForAll{$front_i^n$ in $image_i$}
            \ForAll{$front_{i-1}^k$ in $image_{i-1}$}
                \State $\Delta\epsilon$ = mean($\epsilon_i^n$) - mean($\epsilon_{i-1}^k$)
                \If{$-3.2<=\Delta\epsilon <= 6.7$}
                    \State $front_i^n$ and $front_{i-1}^k$ belong to same CME  \Comment{Min. $\Delta\epsilon$ and  $\Delta PA$ are associated}
                \EndIf
            \EndFor
            \If {$front_i^n$ not linked to previous CME, and flagged as new CME}
                \State $front_i^n$ is new CME
            \Else
                \State $front_i^n$ is discarded
            \EndIf
        \EndFor    
    \ElsIf{count($front_{i}$) = 0 and count($front_{i-1}$) $>$ 0}
        \State CME has ended
    \Else
        \State No CME present
    \EndIf
\end{algorithmic}
\end{algorithm}

\subsection{Evaluation}\label{subsec:evaluation}

Accurately evaluating machine learning models is crucial for determining their effectiveness and utility. Furthermore, it is important to select metrics that are both representative and suitable for the specific task at hand. We aim to evaluate our model's capabilities for CME segmentation and tracking.

For the segmentation task, the STEREO/HI image sequences are randomly distributed between datasets, allocating 70~\% of the data to the training, 20~\% to the test, and 10~\% to the validation set. All images belonging to the same CME are always part of the same set to prevent leakage between them. To provide a more reliable estimate of the model’s generalization ability, we perform 5-fold cross-validation. For cross-validation, the data is divided into five equally sized chunks, so-called folds. The model is then trained on four folds, with a percentage of the training data withheld for validation, while one fold is reserved for testing. This process is repeated five times, using different folds for the training and test sets, yielding five trained models. The final performance metrics are calculated by averaging the results on the test dataset across all five models. On average, our training set contains 5869 images with CMEs, while the validation and test sets consist of 902 and 1693, respectively. The distributions of CME and non-CME images for each of the five models can be found in Table \ref{tab:Data_Distribution_Folds}.

\begin{table}
\caption{Distribution of STEREO/HI images containing/not containing CMEs across the training, test, and validation sets for all five models evaluated as part of the 5-fold cross-validation. We attempt to maintain a similar ratio of CME to non-CME images within each set for all models.}
\centering
\begin{tabular}{l l c c c}
\hline
 Model No. & Class  & Training & Test & Validation  \\
\hline
 Model\_1 & CME  & 6068 & 1858 & 537  \\
         & No CME  & 2718 & 787 & 435  \\
 Model\_2 & CME  & 6039 & 1465 & 959  \\
         & No CME  & 2944 & 766 & 230  \\
 Model\_3 & CME  & 5996 & 1503 & 964  \\
         & No CME  & 2728 & 635 & 577  \\
 Model\_4 & CME  & 5740 & 1711 & 1012  \\
         & No CME  & 2505 & 961 & 474  \\
 Model\_5 & CME  & 5501 & 1926 & 1036  \\
         & No CME  & 2871 & 791 & 278  \\
\hline
\end{tabular}
\label{tab:Data_Distribution_Folds}
\end{table}

To evaluate our model's performance on the segmentation task, we compare the predicted masks to the ground truth on a per-pixel basis, making the definitions of true positives (TPs), false positives (FPs), true negatives (TNs), and false negatives (FNs) straightforward. A pixel classified as part of a CME by the model is a true positive if it also belongs to a CME in the ground truth. If the pixel is instead classified as a CME when it is part of the background, it is a false positive. Conversely, a true negative occurs when a pixel is correctly identified as background, while a false negative arises when a pixel belonging to a CME is misclassified as background. Since the model sees each image multiple times, and thus produces multiple predictions for the same image, the resulting masks must be aggregated to obtain a single final prediction for each image. To derive the final mask, we take the mean, median, and maximum across all predictions for each image. We compare these three aggregation methods to determine the most suitable one for the task. To evaluate segmentation performance, we compute the IoU, precision, recall, and Dice score, which are defined as follows:

\begin{enumerate}
  \item Recall: measures how many of the true positives in the dataset the model managed to identify. It ranges from 0 to 1, with a score of 1 indicating that the model has correctly classified all positive samples in the dataset. Recall is defined as $\text{Recall} = \frac{\text{TP}}{\text{TP} + \text{FN}}$.
  
  \item Precision: measures the model's ability to correctly identify positive instances; in other words, the quality of a positive prediction. Precision also ranges from 0 to 1, and is usually anti-correlated with recall. A precision of 1 would indicate that every positive prediction corresponds to a true positive sample. Precision is defined as $\text{Precision} = \frac{\text{TP}}{\text{TP + FP}}$.
  
  \item Dice coefficient: also known as F1 score. Measures the overlap of the predicted mask and the ground truth. The Dice coefficient is defined as $\text{Dice} = \frac{2 \text{TP}}{2\text{TP} + \text{FP} + \text{FN}}$.
  
  \item Intersection-over-Union (IoU): also known as Jaccard index. It measures the overlap between a predicted mask and its ground truth, and is thus very similar to the Dice coefficient. The IoU penalizes over- and under-segmentation more than the Dice coefficient. The IoU score is defined as $\text{IoU} = \frac{\text{TP}}{\text{TP} + \text{FP} + \text{FN}}$.
\end{enumerate}

To evaluate tracking performance, we employ two different approaches: event-based evaluation and continuous evaluation. For both methods of tracking evaluation, we compute precision and recall, as well as the mean of the absolute error of the tracks’ start and end times, given by:

\[
\Delta t_{\text{start}} = \left|t_{\text{start}}^{\text{groundtruth}} - t_{\text{start}}^{\text{predicted}}\right|
\]

\[
\Delta t_{\text{end}} = \left|t_{\text{end}}^{\text{groundtruth}} - t_{\text{end}}^{\text{predicted}}\right|
\]

For event-based evaluation, we assess the model’s ability to track CMEs by comparing the predicted tracks to the tracks from the ground truth annotations. Event-based evaluation utilizes the same models as the segmentation task and follows the same cross-validation procedure, meaning that the final scores are averaged across all five models.

For this mode of evaluation, CMEs are counted as true positives if they can be matched with a CME in the ground truth. To be considered a potential match, the CME must exhibit both a sufficient temporal and spatial overlap.

Sufficient temporal overlap is defined as:

\[
t_{\text{overlap}} = \frac{\text{overlap}}{\text{duration}} > 0.25
\]

where ``overlap'' is the shared time between both CMEs, and ``duration'' is the total duration of both CMEs combined. Sufficient spatial overlap is given if $\text{IoU} > 0.1$.

Each predicted CME can only be matched with at most one CME from the ground truth, and vice versa. It is possible for a CME from either dataset to have no match in the other. To enforce this, only the predicted CME with the smallest absolute time error is associated with a given ground truth CME. The absolute time error is given by:

\[
\Delta t_{\text{total}} = \Delta t_{\text{start}} + \Delta t_{\text{end}}
\]

A CME is considered a false positive if it cannot be associated with any CME in the ground truth. Conversely, a false negative occurs when a ground truth CME cannot be associated with any predicted CME.

For continuous tracking evaluation, we aim to assess the model’s performance in a more realistic scenario, where it processes a continuous data stream rather than isolated event sequences. To facilitate this, we train an additional model identical to the previous ones in architecture, but with a different data split. Specifically, this model is trained on 90~\% of the available data and validated on the remaining 10~\%, without setting aside a dedicated test dataset. This model is then applied to a continuous two-year time series of images and compared against CME tracks from the HIGeoCAT catalog.

To capture different solar activity conditions, we select the years 2009 and 2011 for continuous tracking evaluation, corresponding to solar minimum and the rising phase of solar cycle 24, respectively. While the official solar maximum of cycle 24 occurred in late 2014, STEREO-A was approaching superior solar conjunction at that time, making data interpretation more challenging due to increasing observational gaps and reduced data quality. As a compromise between solar activity levels and spacecraft viewing geometry, 2011 was chosen to represent an active period with favorable observational conditions.

The definitions of TPs, FPs, and FNs closely follow that of event-based evaluation, with HELCATS tracks being used as the ground truth. Some modifications are necessary in the definition of a potential match, as there are no ground truth segmentation masks to compare to (and thus no IoU score to compute) and HELCATS tracks are only defined along a single PA, meaning that we can't rely on spatial overlap of fronts as a criterion. Temporal overlap is not affected by this, but the definition of spatial overlap must be changed to reflect these factors. We define the spatial overlap as the mean difference in $\epsilon$ along each PA for the predicted and ground truth CME fronts:

\[
\Delta \epsilon = \operatorname{mean}\left(\left| \epsilon_{\text{pa}}^{\text{groundtruth}} - \epsilon_{\text{pa}}^{\text{predicted}} \right|\right).
\]

To ensure that no CME in the predicted dataset is matched to more than one event in the ground truth dataset, and vice versa, we assign the TP label for a given ground truth CME to the predicted CME with the smallest combined $\Delta t_{\text{total}}$ and $\Delta\epsilon$. The definition of FPs and FNs remains analogous to that introduced for the event-based tracking evaluation.

\section{Results}
\label{sec:results}

\subsection{Segmentation}

\begin{figure}
\includegraphics[width=\textwidth]{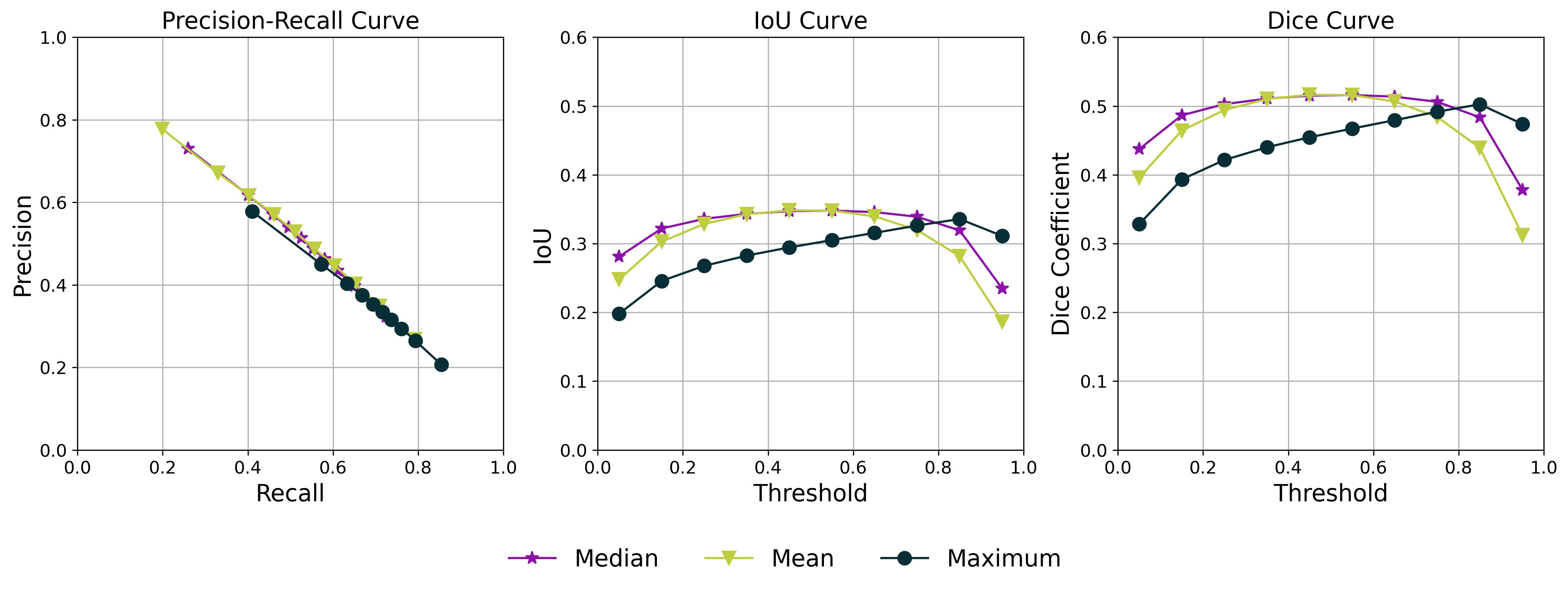}
\caption{Key metrics for segmentation performance achieved using different aggregation methods (colored), averaged for all 5 models. The leftmost panel shows precision plotted against recall. The middle panel and rightmost panel display the IoU and Dice Score, respectively, at various binarization thresholds ranging from 0.05 to 0.95.}
\label{fig:curves_reduction_techniques}
\end{figure}

Key segmentation performance metrics are presented in Figure \ref{fig:curves_reduction_techniques} for different aggregation methods, evaluated across binarization thresholds ranging from 0.05 to 0.95. The results indicate a very similar overall performance across aggregation methods and thresholds. As expected, the maximum method performs better in terms of Dice score and IoU at higher binarization thresholds compared to the mean and median methods. Visually, the precision-recall curves presented in the leftmost panel of Figure \ref{fig:curves_reduction_techniques} look alike for the mean and median methods, while the range values for both precision and recall across different thresholds are considerably smaller for the maximum method.

We identify the optimal threshold, as defined by the highest IoU, for each aggregation method and summarize the metrics at this threshold in Table \ref{tab:segmentation}. While the optimal threshold varies between methods, with threshold(median) = 0.55, threshold(mean) = 0.45, and threshold(max) = 0.85, the resulting metrics don't exhibit significant variations. The best IoU achieved varies between 0.34 for the maximum method and 0.35 for the median method and mean methods. The Dice score exhibits similar variation, with both the mean and median methods scoring at 0.52, while the maximum method achieves a score of 0.50. The only notable differences lie with the precision and recall scores. In terms of precision, the maximum method performs worse than the mean and median methods with a score of 0.45, while the other two methods achieve scores of 0.49 and 0.51, respectively. Since precision and recall are usually anti-correlated, the opposite is true for the recall metric, with the maximum method scoring 0.57, while the mean and median methods achieve scores of 0.56 and 0.53, respectively. We find that our model is largely insensitive to the choice of aggregation method and threshold, and select the mean method at threshold(mean) = 0.45 as the optimal method-threshold combination since it reaches the highest combined metrics out of all method-threshold combinations.
 
\begin{table}\label{tab:segmentation}
\caption{Key metrics for different aggregation methods, averaged across all five models, given at the binarization threshold value that results in the highest IoU for the respective method.}
\centering
\begin{tabular}{l c c c c c}
\hline
 Method  & Threshold & Precision & Recall & IoU & Dice \\
\hline
  Median & 0.55 & 0.51 & 0.53 & 0.35 & 0.52 \\
  Mean & 0.45 & 0.49 & 0.56 & 0.35 & 0.52 \\
  Maximum & 0.85 & 0.45 & 0.57 & 0.34 & 0.50 \\
\hline
\end{tabular}
\end{table}

\begin{figure}
\includegraphics[width=\textwidth]{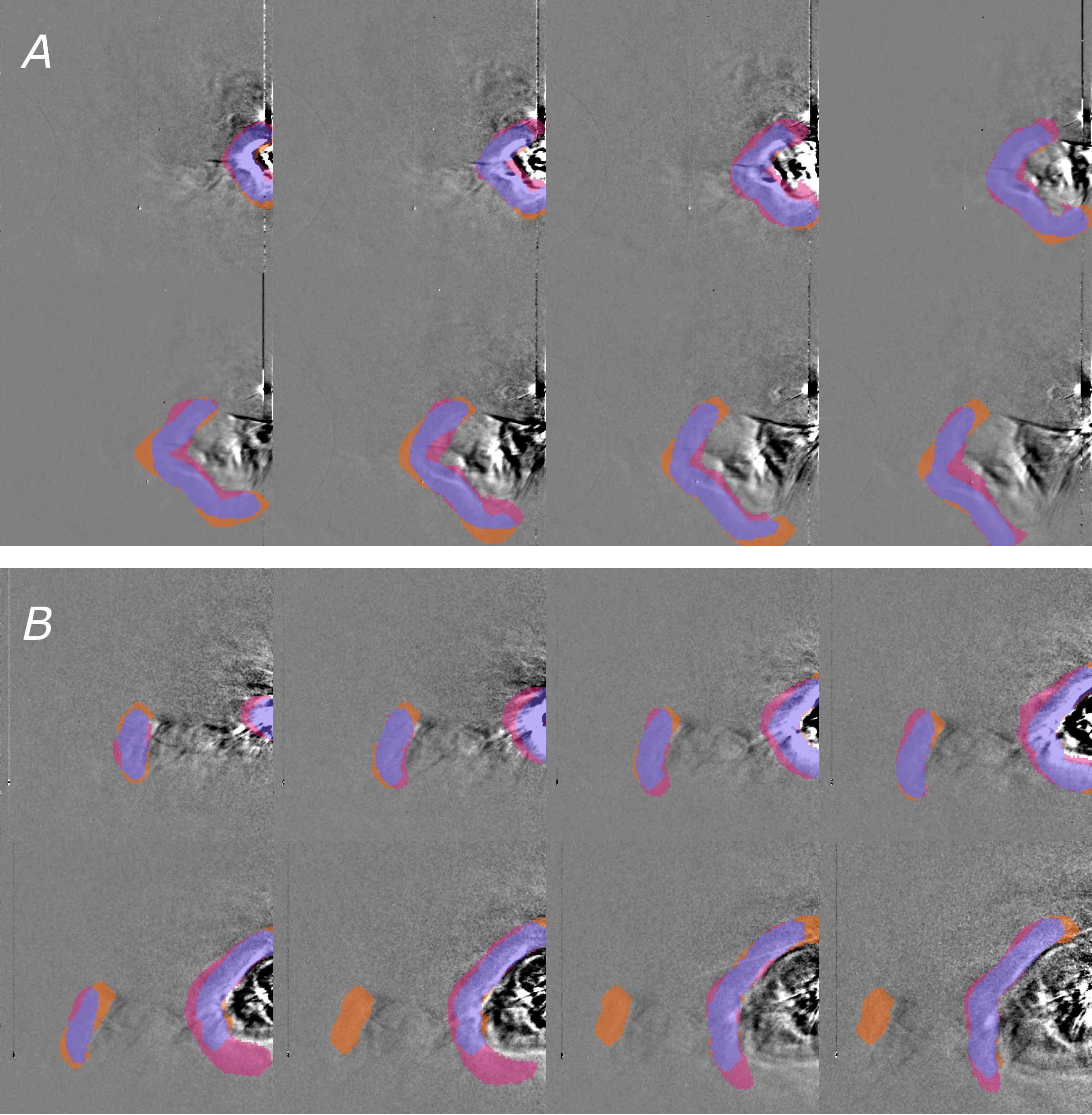}
\caption{Sequences of images showcasing the model's segmentation performance using the best method-threshold combination threshold(mean) = 0.45. The overlap between the predicted front and the ground truth mask is shown. Each pixel belonging to the front is classified as a TP (purple), FP (pink), or FN (orange). Panel (A) shows a single CME front close to the Sun, while panel (B) shows two CME fronts, one being further away from the Sun and thus fainter than the other one.}
\label{fig:3colorplot}
\end{figure}

Figure \ref{fig:3colorplot} shows example segmentation results for post-processing using threshold(mean) = 0.45. The upper panel showcases a well-segmented CME, located at the beginning of the HI1 field of view. CMEs close to the Sun tend to be brighter and more easily distinguishable from the background. The model captures most of the front with only minor discrepancies at the edges compared to the ground truth. This example also highlights that an inherent challenge in CME segmentation lies in defining the precise boundary of a CME front, which is subjective. Even minor deviations of the predicted from the ground truth mask negatively impact performance metrics despite the segmentation being visually accurate. The lower panel of Figure \ref{fig:3colorplot} presents a more challenging case. There are two CMEs in the image, with one being further away and thus slightly fainter than the other CME. Our model accurately segments the front of the CME closest to the Sun, but struggles to capture the complete front of the preceding CME as it moves outwards.

\subsection{Event-based Tracking}

Table \ref{tab:tracking_ml} summarizes the tracking performance for event-based evaluation, analyzed using the optimal method-threshold combination of threshold(mean) = 0.45. Compared to the ground truth tracks, the model achieves an average $\Delta t$ of 1.42 hours for CME start times and 4.71 hours for CME end times. This difference in results between the average $\Delta t_\text{start}$ and $\Delta t_\text{end}$ reflects the model’s difficulty in tracking faint CME fronts, which often causes the track to be cut short. The model successfully recovers 145 of the 257 ground truth tracks, leaving 112 unmatched, and additionally identifies 23 tracks not present in the ground truth catalog. This corresponds to a precision of 0.87 and a recall of 0.56. The results are consistent across all five models, with standard deviations of 0.41 hours and 0.76 hours for the $\Delta t_\text{start}$ and $\Delta t_\text{end}$, respectively. Precision and recall vary by 0.06, and 0.05 respectively, indicating stable performance and good generalizability across different data splits.

\begin{table}
\caption{Summary of key metrics used to gauge event-based tracking performance. The best method-threshold combination threshold(mean) = 0.45 was used to generate the results. Results for average absolute start ($\Delta t_\text{start}$) and end ($\Delta t_\text{end}$) time error, as well as precision, and recall, were averaged across all 5 models. Results for true positives, false positives, and false negatives were summed up instead to allow for easier comparison to the total number of ground truth CMEs in the given time frames.}
\centering
\begin{tabular}{l r}
\hline
 Metric  & Result \\
\hline
  $\Delta t_\text{start}$ & 1.42 h \\
  $\Delta t_\text{end}$ & 4.71 h \\
  Precision & 0.87 \\
  Recall & 0.56 \\
  Total CMEs & 257 \\
  True Positives & 145 \\
  False Positives & 23 \\
  False Negatives & 112 \\
\hline
\end{tabular}
\label{tab:tracking_ml}
\end{table}

\begin{figure}
\includegraphics[width=\textwidth]{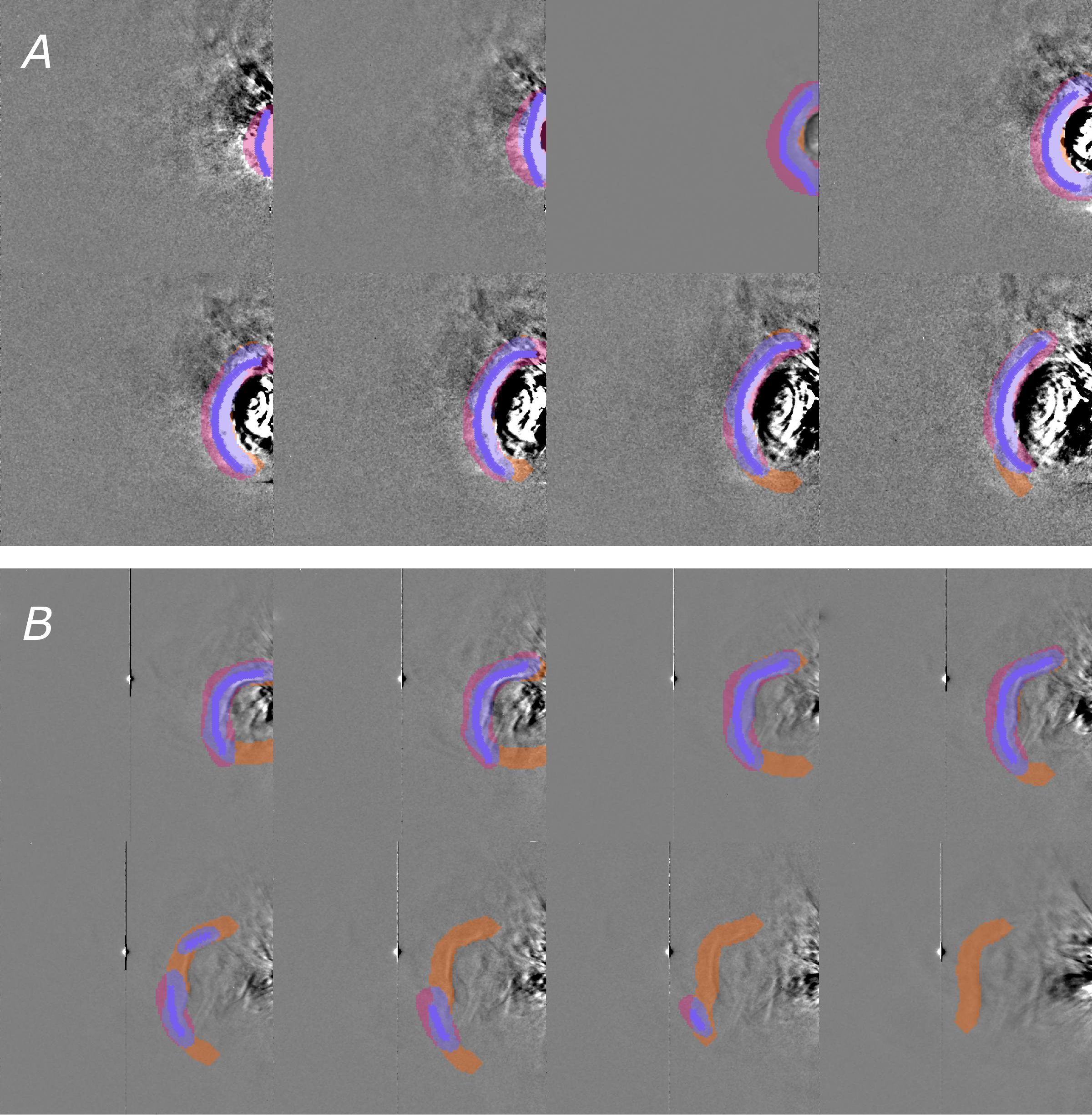}
\caption{Time-series' of STEREO/HI data with results from event-based tracking overlaid. Pixels are classified into true positives, false positives, and false negatives, and subsequently color-coded to show segmentation results as in Figure \ref{fig:3colorplot}. The fronts identified by the tracking algorithm are overlaid in purple, indicating that they were correctly identified as belonging to a ground truth CME. Panel (A) demonstrates the models' ability to accurately segment and track CMEs close to the Sun, while Panel (B) shows an example of decreasing performance in both tasks as the CME grows fainter.}
\label{fig:tracking_ml}
\end{figure}

Figure \ref{fig:tracking_ml} provides qualitative examples of the tracking algorithm’s performance. The upper panel shows a case where the model successfully detects and tracks a bright CME front relatively close to the Sun, maintaining good performance throughout the sequence. In contrast, the lower panel depicts a CME that is slightly fainter, with tracking performance gradually degrading as the structure propagates outward. These examples highlight both the model’s strengths in identifying clear, well-defined CME fronts and its limitations when dealing with weaker, more ambiguous signals.

\subsection{Continuous Tracking}

Table \ref{tab:tracking_helcats} summarizes the results of continuous tracking for both years under study. As in the case of event-based tracking, the average $\Delta t_{\text{end}}$ is notably higher than the average $\Delta t_{\text{start}}$: 6.84 hours (2009) and 6.97 hours (2011) for end times, compared to 3.20 hours (2009) and 3.30 hours (2011) for start times. Analyzing the remaining metrics reveals substantial differences between the two years. Precision dropped from 0.76 in 2009 to 0.56 in 2011. Recall is higher for 2011 (0.85) than for 2009 (0.64).

These differences are closely linked to the solar cycle. In 2011, closer to the solar maximum, the HIGeoCAT catalog contains a significantly higher number of CMEs (116) than in 2009 (55). While the number of false negatives remains almost unchanged between the two years, the number of false positives increases sharply in 2011, with 78 FP tracks compared to just 11 in 2009. This suggests that the model struggles more in periods of heightened solar activity, likely due to the increased complexity and frequency of overlapping CME events.

\begin{table}
\caption{Summary of key metrics used to gauge continuous tracking performance. The best method-threshold combination threshold(mean) = 0.45 was used to generate these results. Metrics are listed separately for 2009 and 2011.}
\centering
\begin{tabular}{l r r}
\hline
 Metric  & 2009 & 2011 \\
\hline
  $\Delta t_{\text{start}}$ & 3.20 h & 3.30 h \\
  $\Delta t_{\text{end}}$ & 6.84 h & 6.97 h \\
  Precision & 0.76 & 0.56 \\
  Recall & 0.64 & 0.85 \\
  Total CMEs & 55 & 116 \\
  True Positives & 35 & 99 \\
  False Positives & 11 & 78 \\
  False Negatives & 20 & 17 \\
\hline
\end{tabular}
\label{tab:tracking_helcats}
\end{table}

\begin{figure}
\includegraphics[width=\textwidth]{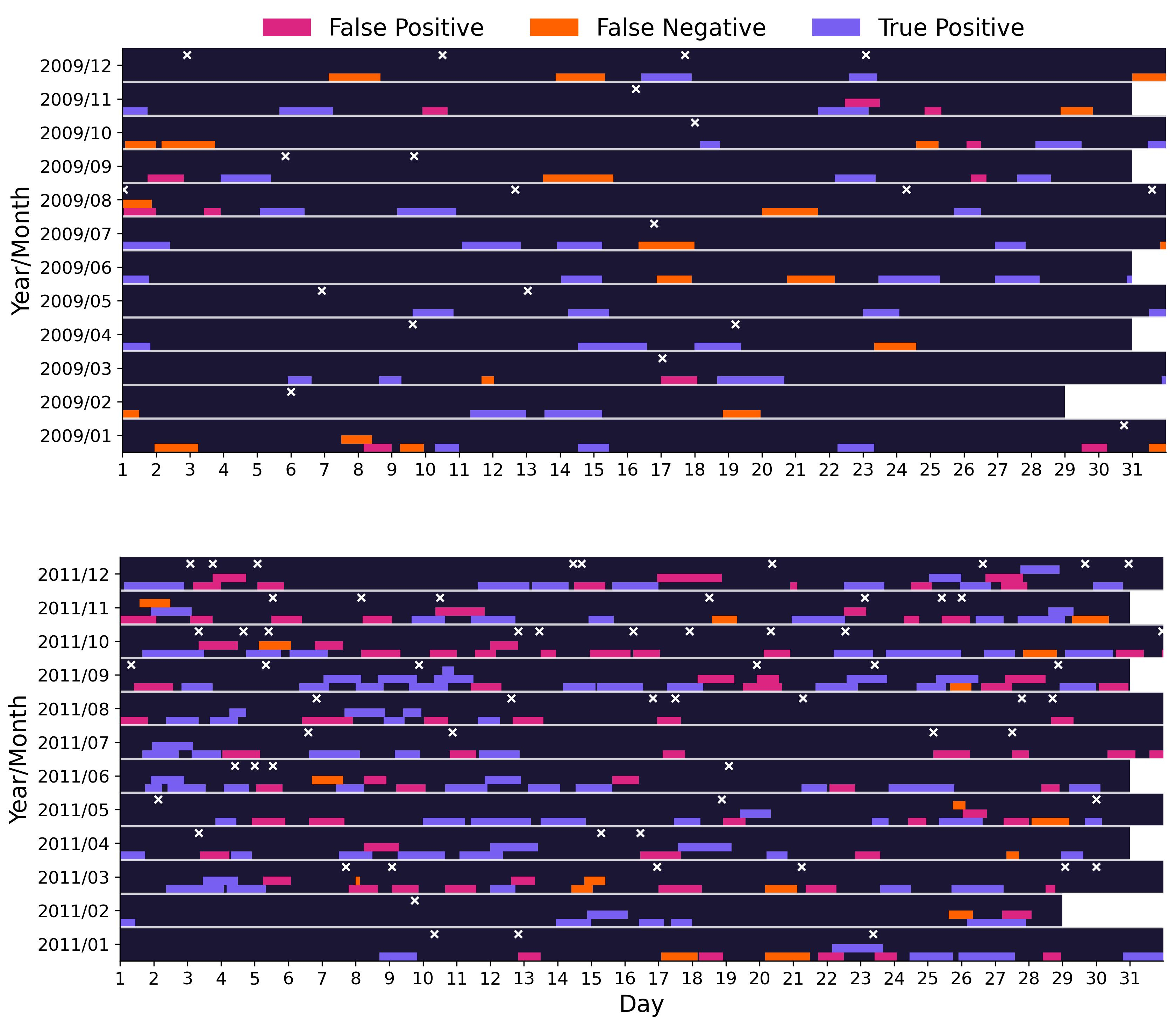}
\caption{Results of continuous tracking evaluation for the year 2009 (top panel) and 2011 (bottom panel). Tracks obtained using our algorithm are compared to HIGeoCAT tracks of the same time period and assigned either a TP (purple) or an FP (pink) label. The beginning and end times of our track along the PA given by the corresponding HELCATS track are shown in the case of a TP detection. If a track is labeled as an FP, the minimum beginning time and maximum end times across all PAs are plotted. HIGeoCAT tracks that could not be associated with any of our tracks are labeled as FN (orange), with the beginning and end times as provided in HIGeoCAT. Furthermore, the starting times of CMEs identified as part of HICAT, but not tracked as part of HIGeoCAT, are marked with white crosses. Each row corresponds to a month within the year. The vertical positions of the tracks within a given month were chosen to prevent overlap during plotting.}
\label{fig:heatmap_plot}
\end{figure}

Figure \ref{fig:heatmap_plot} provides a visualization of the number of true positive, false positive, and false negative detections. The overall increase in CME activity from 2009 to 2011 is immediately apparent, as is the substantial rise in false positive detections during the more active year. Closer examination reveals that several CMEs listed in the HIGeoCAT catalog are missed by the model, resulting in false negative cases. However, in many of these instances, the model does detect a CME in the relevant time frame but fails to track it consistently along the HELCATS-defined position angle. This indicates that while the front is often identified, incomplete segmentation of the front can lead to fragmented or misaligned tracks that do not meet the matching criteria.

It is important to note that only CMEs classified as fair or good according to the HELCATS criteria are included in HIGeoCAT, while poor-quality events are excluded. Furthermore, CMEs at very high or low position angles are often not tracked due to the limited field of view of the HI instruments. Events at both very high or low latitudes become more common near solar maximum. This affects the completeness of HIGeoCAT, and thus the model evaluation based on it. Therefore, any apparent decrease in model performance during solar maximum should be interpreted in the context of these inherent limitations in the reference data.

HICAT includes additional events not present in HIGeoCAT, providing their starting times within the HI1 FOV, but no time-elongation tracks. This adds 20 CMEs for 2009, all of which are classified as poor, and 63 for 2011. Of the 63 additional CMEs in 2011, 11 are classified as good, 32 as fair, and 20 as poor. To account for these CMEs, we attempt to associate unmatched predictions from our model with the HICAT CMEs using a simple time-based matching criterion. Since HICAT only provides a starting time, we rely on temporal proximity alone. As a starting point, we consider the mean starting time error $\Delta t_{\text{start}}$ between our model and HIGeoCAT CMEs, which is $\pm 3.20$ hours for 2009 and $\pm 3.30$ hours for 2011. Rounding these values to the nearest integer multiple of the HI-1 image cadence (40 minutes) yields a matching window of $\pm 3.33$ hours, which we adopt as a our threshold. Applying this criterion yields 2 additional true positives for 2009 and 30 for 2011.

\begin{figure}
\includegraphics[width=\textwidth]{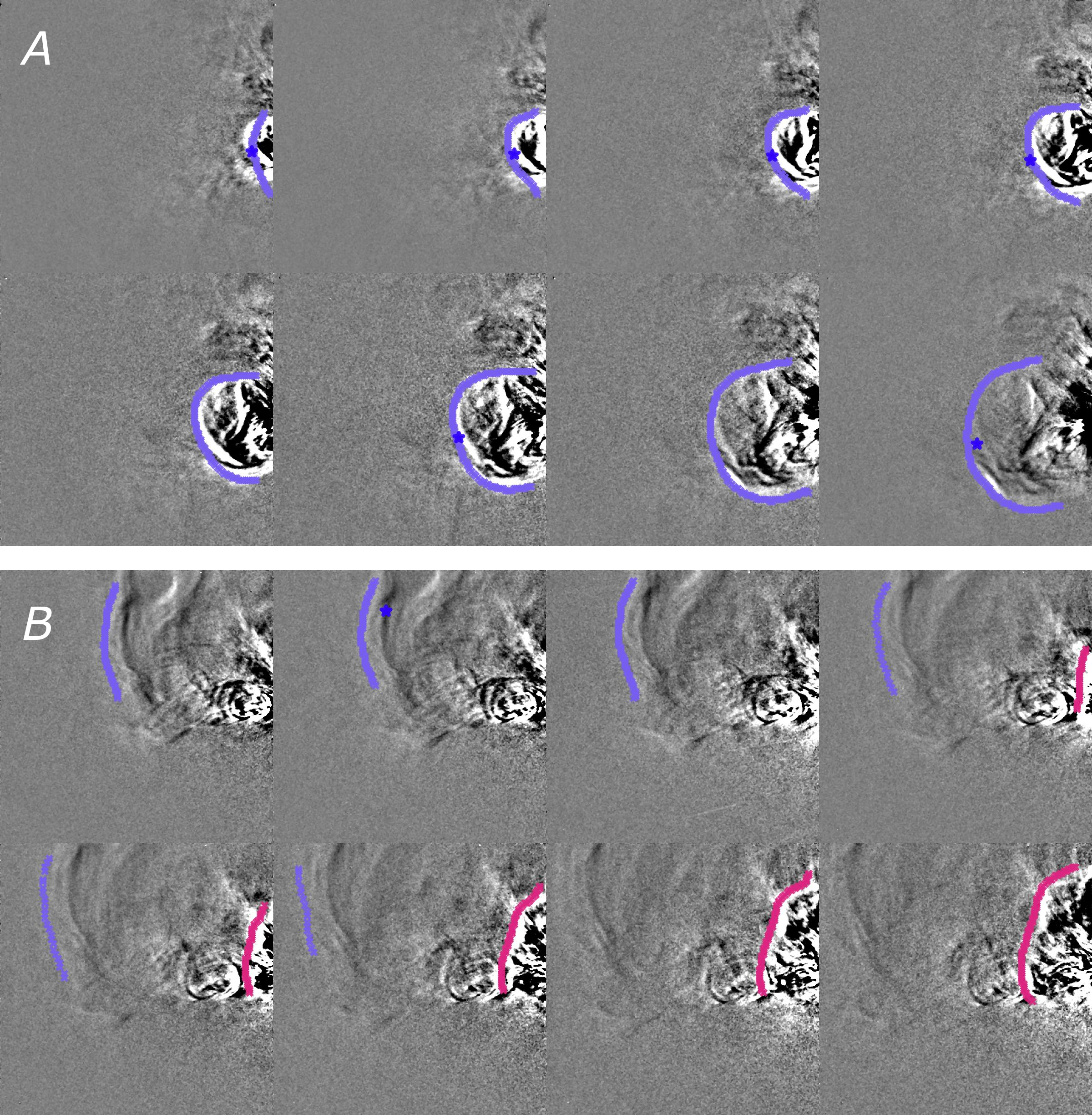}
\caption{Examples of STEREO/HI image sequences showcasing the performance of our algorithm for continuous tracking. Tracks that were successfully associated with a HIGeoCAT track are marked in purple, while ones for which this is not the case are marked in pink. The corresponding HIGeoCAT track is marked with a dark-purple star, where applicable. Note that some frames do not contain any HIGeoCAT data points as tracking is done in Jmaps, not images, resulting in skipped images during tracking. Panel (A) shows an example of a well defined, singular CME, resulting in a detected front that is clean and cohesive. Panel (B) shows an example of a CME immediately followed by a smaller structure, possibly the core, which is then identified as the beginning of a new CME. Despite the fact that this structure is likely to be part of the preceding CME, identifying it as a separate event could potentially be useful from an operational perspective, as both the front and core of a CME can be geoeffective.}
\label{fig:tracking_helcats}
\end{figure}

Figure \ref{fig:tracking_helcats} offers a direct visual comparison between the CME fronts detected and tracked by our model and those from the HIGeoCAT catalog. As in previous examples, the model often struggles to consistently identify the correct front when multiple structures are present in the image. Nevertheless, in both cases, at least one of the CMEs could still be matched to a HIGeoCAT track. A closer comparison between the HIGeoCAT annotations and the model outputs reveals that different parts of the same CME are sometimes being tracked. While our model tends to follow the outermost edge of the front, the HELCATS catalog occasionally traces structures located slightly further behind. This discrepancy underscores the inherent challenge of precisely defining the spatial extent of a CME front, especially in complex or faint events.

\section{Conclusion and Discussion}

The difficulty of consistently interpreting white-light images, as well as their increasing availability, have sparked the need for efficient, automated CME detection and tracking methods. In this work, we present STRUDL, a model capable of detecting CMEs in STEREO/HI data. By using sequences of images instead of individual frames as input for our model, we introduce temporal context to improve the consistency of the segmentation. Our model performs well in identifying bright CMEs but struggles with fainter and more diffuse fronts, especially near the edges of the HI1 FOV. The challenge of CME segmentation is exacerbated by the inherent ambiguity in defining CME fronts. Small mismatches between predicted and ground-truth masks can lead to a significant drop in performance metrics, such as IoU, even when the segmentation is visually accurate.

We also implement a basic tracking algorithm, which works with output from STRUDL and is able to consistently link CMEs detected by our model across images. We evaluate our tracking algorithm by comparing to our own ground-truth tracks (event-based evaluation), as well as to CMEs tracked as part of the HIGeoCAT catalog (continuous evaluation). In both cases, we observe a higher mean absolute error for CME end times compared to start times, again indicating that the model struggles to fully track events through the HI1 FOV. The model produces several false positive tracks, particularly during solar maximum. Some of these tracks may correspond to real CMEs that were not present in the ground truth datasets. This hypothesis appears to at least partially explain the high number of false positives in the case of continuous tracking evaluation. Not all CMEs during the given periods were tracked in HIGeoCAT, and incorporating additional HICAT detections into our evaluation shows that a portion of our model's apparent false positives detections are associated with real CMEs in the HICAT catalog. Inspecting the quality flags of the unmatched HIGeoCAT CMEs reveals that most of the CMEs not detected by our model are of fair quality (16 out of 20 for 2009, 13 out of 17 for 2011), while most CMEs flagged as good quality were correctly identified. Overall, the model can identify and track most CMEs present in the ground truth datasets, proving its value as a potential early-warning tool in the future.

Improvements to both segmentation and tracking performance could be made by a different approach to CME annotation. Our current annotation approach marks only the leading edge of each CME, which provides a consistent but limited representation of CME structure. As a result, the model learns to focus on identifying bright fronts, potentially leading it to identify other bright structures (e.g. the core) as new CME fronts. This is particularly apparent in images where multiple CMEs overlap or interact with each other. Distinguishing between successive CMEs and complex internal structures is important from a scientific perspective, and using a more comprehensive annotation strategy, such as using polygons to outline the entirety of a CME, could provide the model with more contextual information. From an operational perspective, however, this distinction is less significant. Both the front and the core of a CME have the potential to cause disruptions at Earth; being alerted to and tracking both separately could thus be a desirable outcome. Furthermore, while the current tracking implementation is based on simple heuristics, more advanced approaches, such as data-driven (e.g. \citeA{liObject2022}) or probabilistic models (e.g. \citeA{spilgerDeep2021}), could provide more coherent and robust tracking. These techniques may also enable estimation of CME parameters, allowing us to predict CME arrival time and speed, thus opening up more possibilities for assessing performance.

Currently, the model is trained and evaluated on STEREO/HI science data. For real-time applications, however, the model must be able to operate on low-latency, low-resolution beacon data. To facilitate this, enhancements to the input data may be necessary. \citeA{lelouedecBeacon2Science2025} have developed the Beacon2Science model, a machine-learning algorithm trained to convert STEREO/HI beacon data into enhanced beacon data, which has a spatial and temporal resolution closer to that of science data.

To speed up the generation of beacon training data and reduce the time spent on manual annotation, our pre-trained model may be used in a semi-automatic workflow where predicted masks are manually refined and projected onto beacon data. This process could significantly reduce the effort required compared to fully manual annotation, not only for generating beacon training data, but also for increasing the amount of available science training data. This is particularly relevant for preparing for future missions that will carry HI instruments, such as ESA's Vigil mission, scheduled for launch in 2031, or for utilizing data from the PUNCH mission, which is already operational. Further improvements in both segmentation and tracking are needed to ensure reliable performance across the full range of solar conditions. Overall, our model provides a promising foundation for automated CME detection and tracking in HI data.

\section*{Open Research Section}\label{sec:openresearch}

The STEREO/HI data used in this paper can be downloaded from \url{https://stereo-ssc.nascom.nasa.gov/pub/}. We provide the code for downloading and processing the data in \citeA{STEREOHIDataProcessingv1.1.0}. The code to run STRUDL is available in \citeA{STRUDLv1.0.0}. The tracking tool is available as part of the same repository. The HICAT and HIGeoCAT catalogs are available under \url{https://www.helcats-fp7.eu/catalogues/wp2_cat.html} and \url{https://www.helcats-fp7.eu/catalogues/wp3_cat.html}, respectively. Data from the Solar Stormwatch II citizen science project was used in the creation of this manuscript \cite{scottSolar2025}. The annotations used to train STRUDL are available on Figshare \cite{bauerAnnotations2025}. The configuration files needed to reproduce model results, as well as the saved weights of all of our models are available on Figshare \cite{bauerSTRUDL2025}. This research makes use of the AstroPy \cite{collaborationAstropy2022} and Pytorch \cite{paszkePyTorch2019} libraries for Python.

\acknowledgments
This research was funded in whole or in part by the Austrian Science Fund (FWF) [10.55776/P36093]. For open access purposes, the author has applied a CC BY public copyright license to any author-accepted manuscript version arising from this submission. This work is supported by ERC grant (HELIO4CAST, 10.3030/101042188) funded by the European Union. Views and opinions expressed are however those of the author(s) only and do not necessarily reflect those of the European Union or the European Research Council Executive Agency. Neither the European Union nor the granting authority can be held responsible for them. DB recognises the support of the UK Space Agency for funding STEREO/HI operations in the UK. This publication uses data generated via the Zooniverse.org platform, development of which is funded by generous support, including a Global Impact Award from Google, and by a grant from the Alfred P. Sloan Foundation. LB is supported by a UKRI Future Leaders Fellowship (DARES, MR/Y021207/1).

%
%
\clearpage
\bibliography{mybib}

%
%
%
%
%

\end{document}